\documentclass{aa}  
\usepackage{graphicx}
\usepackage{txfonts}
\begin{document}
   \title{Probable nonradial $g$-mode pulsation in early A-type stars\thanks{Based on
       observations collected at ESO-La Silla; program 75.D-0342}}

   \author{E. Antonello \inst{1}
          \and
          L. Mantegazza \inst{1}
	  \and 
  	  M. Rainer \inst{1}
          \and
          A. Miglio \inst{2}
          }

   \offprints{E. Antonello}

   \institute{INAF-Osservatorio Astronomico di Brera, Via E. Bianchi
     46, 23807 Merate, Italy\\
     \email{elio, luciano, rainer@merate.mi.astro.it}
        \and
     Institut d'Astrophysique et de G\'eophysique de l'Universit\'e de
     Li\`ege, 17 All\'ee du 6 Ao\^ut, 4000 Li\`ege, Belgium\\
      \email{miglio@astro.ulg.ac.be}
             }

   \date{  ; }

   \abstract
   {A survey for line profile variability in early 
   A-type stars has been performed in order to detect
   nonradial pulsation signatures. The star HR 6139, with 
   spectral type A2V and estimated $T_{\rm{eff}}=8800$ K,
   shows evident line profile variations that
   can be explained by oscillations in prograde $g$-modes. This feature and
   the known photometric variability are similar to those
   observed in the Slowly Pulsating B-type stars. However HR 6139
   is much cooler than the cool border of the instability strip of
   such variables, and it is hotter than the blue edge of the
   $\delta$ Scuti instability strip. There are indications of a tiny 
   variability also in other four objects, whose nature is not yet 
   clear.

   }

   \maketitle
%
%________________________________________________________________

\section{Introduction}

The modelling of convection is one of the main astrophysical problems that are 
to be solved, and the proposed theories cannot be adequately tested 
since an effective experimental tool is still lacking. 
The new satellite COROT (Baglin et al.,
\cite{bag}), that will be launched in 2006, is intended to 
probe stellar interiors by means of asteroseismology, 
though only the modes with lowest degree $\ell$ can be detected 
photometrically. 
Asteroseismology promises to be a very effective tool to obtain detailed 
and independent information on the size of convective zones and on the 
extent of overshooting.
An analysis of solar-like $p$-modes of main sequence models shows that, 
whereas the seismic signatures of the convective envelope are very small 
(e.g. Ballot et al., \cite{bal}), larger effects are predicted for convective 
cores of intermediate-mass stars (e.g. Roxburgh \& Vorontsov, \cite{rv}; 
Miglio \& Antonello, \cite{mig2}).
As shown by Straka et al. (\cite{stra}), in addition to pressure modes, further 
and more detailed information on the stellar core can be gathered from 
the periods of gravity modes, which are direct probes of the deep 
stellar interior.
It is therefore important to study the seismology of early-type stars 
and, in particular, to detect $g$-mode oscillations.

Known $g$-mode pulsators close to the main sequence include the 
Slowly Pulsating B-type stars (SPB) and the $\gamma$ Dor (F-type) stars.
Pamyatnykh (\cite{pam}) discussed the pulsational domains in the upper 
main sequence, and excluded the presence of pulsators in the spectral 
range B8-A2, i.e. later than the SPB domain and earlier than the
$\delta$ Scuti star domain; however, he remarked that stellar rotation and metallicity
can have an effect on such domains. Townsend (\cite{tow1},
\cite{tow2}) also discussed some effective mechanisms 
that can change the extent of the theoretical instability strip of SPB
stars. 

The finding 
from theoretical three-dimensional simulations of persisting, global $g$-mode 
oscillations (resonances) induced by the convection coupled to
rotation in the early A-type star cores (Browning et al. \cite{bro}) 
opens a new possibility. These modes with low $\ell$ should 
have a frequency of the order of or less than two times the rotation 
frequency. Browning et al. (\cite{bro}), however, give warning on the 
simplifications adopted in their study, and moreover it is not 
clear whether such modes can propagate in the envelope. On the other hand, 
Koen (\cite{koe}), from the analysis of the variable stars 
in the HIPPARCOS catalog, has suggested the existence of a 
possible group of slowly pulsating stars among the early A-type stars.

In the present paper we report the main results of an exploratory
program dedicated to the detection of nonradial mode signatures
in moderately rotating early A-type stars.

%__________________________________________________________________

\section{Observations and data analysis}
The observations were performed with the FEROS spectrograph attached
to the 2.2 m telescope at ESO-La Silla, with resolution 48000 and
spectral range $\lambda$350-930 nm, during a three-night run 
(May, 25-29, 2005). Other spectra were also gathered in June, 18-21,
2005, with the same instrumentation. The targets were selected among
the brightest A0-A3 type stars, with projected rotational velocities
between 50 and 100 km/s; the known binaries and chemically peculiar
stars were excluded. The brightness requirement is due to the very high
$S/N$ ratio needed. The stars were observed cyclically during all the
nights; the typical separation between the spectra of the same 
star was about one hour. The exposure times were estimated in such
a way to get $S/N \ga 300$ in the best spectral range. The nine stars 
are listed in Table 1, along with their apparent
magnitude, the spectral type, the number of spectrograms, the $S/N$ ratio,
the rotational velocity estimated by us, the ratio of
the average standard deviations in the line region and in the 
continuum region with the corresponding $t$-test (discussed below), 
and the results of the survey for variability. In the literature
the MK classification of the nine stars is rather consistent, and
there are no significant discrepancies or indications of peculiarities. 
Furthermore, for two stars, HR 5670 and HR 7950, there are indications
of normal abundances (Andrievsky et al., \cite{and}; Kocer et
al., \cite{koc}). 
\begin{table*}
\caption{Target stars}             
\label{table:1}      
\centering                          
\begin{tabular}{c c c c c r r r c}        
\hline\hline                 
Name (HR) & V &Sp.T. & \# spectr.& $S/N$ & $\upsilon$sin$i$ & s.d. ratio & $t-$test & result \\
\hline                         
5028 & 2.75 & A2V & 26 & 5800 & 88  &  1.310 &  5.40 & var? \\
5332 & 5.53 & A1V & 23 & 2100 & 111 &  1.015 &  0.15 & const \\
5670 & 4.07 & A3V & 23 & 4300 & 69  &  1.099 &  1.59 & const \\
5859 & 5.58 & A0V & 14 & 4100 & 69  &  1.384 &  4.72 & var? \\
5905 & 5.76 & A3V & 24 & 4000 & 89  &  1.204 &  3.59 & var? \\
5925 & 5.12 & A3V & 11 & 5000 & 53  &  2.025 &  7.34 & spectr. bin.   \\
6139 & 6.04 & A2V & 37 & 2900 & 94  &  5.540 & 19.53 & $g$-modes \\
6681 & 5.93 & A0V & 17 & 3400 & 60  & 16.333 &  8.06 & spectr. bin.  \\
7950 & 3.77 & A1V & 10 & 5400 & 112 &  1.219 &  2.82 & var? \\
\hline                          
\end{tabular}
\end{table*}
%-

The spectrograms were reduced and normalized to the continuum level by
means of a pipeline developed by us. We constructed a mean
photospheric profile, using the information contained in
all photospheric lines present in the spectrum. The Least
Squares Deconvolution (LSD) method developed by Donati et
al. (\cite{dona}) consists of deconvolving the observed spectrum
with a mask function including all expected spectral lines with their 
expected depth, as calculated from a model atmosphere, with the 
exclusion of the strongest lines whose profiles are dominated by 
other broadening mechanisms than the rotation. This allows 
to construct a mean line profile combining the information 
contained in thousands of lines, and therefore with 
a very high $S/N$ ratio, typically of several thousands. 
The input parameters for the model atmospheres of our stars were 
estimated using the $uvby{\beta}$ and Geneva photometry data taken 
from SIMBAD database, and with the calibrations of Napiwotzki 
et al. (\cite{napi}) and Kunzli et al. (\cite{kunz}). 
As a by-product we estimated the $\upsilon$sin$i$ values
on the basis of the position of the first zeros of the Fourier
transform of the mean line profile (e.g. Royer et al., \cite{roy}). 
Our $\upsilon$sin$i$ values are not very different from those 
previously known, except for the star HR 5925 (resolved spectroscopic
binary). The uncertainty in the
$\upsilon$sin$i$ values reported in Table 1 is of the order of 1 - 2 km/s.
%-------------------------------------------------------------
   \begin{figure}
   \centering
   \includegraphics[width=9cm]{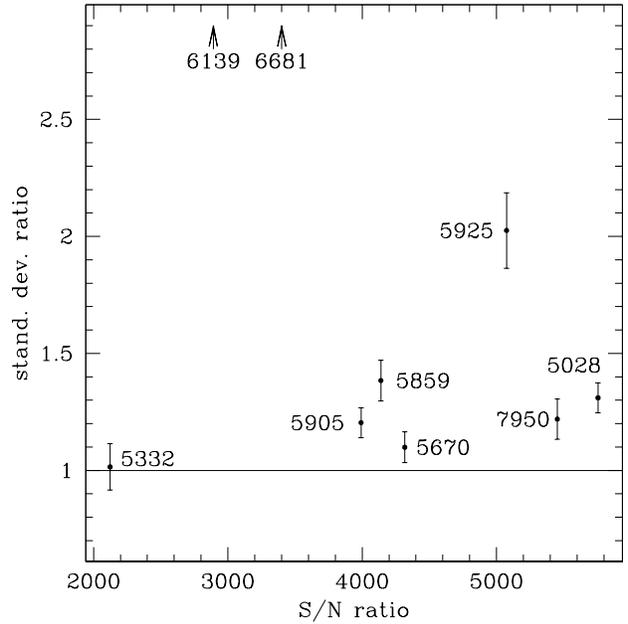}
      \caption{Ratio between the average standard deviations in the 
      line region and in the continuum region, against the $S/N$ ratio.
              }
         \label{Fig1}
   \end{figure}
%-------------------------------------------------------------

We computed the average value of the mean line profile of all
the spectra of a star and the standard deviation in each wavelength bin. The
mean standard deviation in the continuum region allows us to estimate
the mean $S/N$ ratios of the individual mean profiles. In our cases
we estimated $S/N$ values between 2100 and 5800. The comparison 
of the standard deviation values inside and outisde the line profile
allows us to detect the possible variations of the line profile
shape. From a visual examination of these curves, three stars showed
evident variability: HR 6139, HR 5925 and HR 6681. However, in order
to give a quantitative estimate of the variability, two average
standard deviations were computed: one in the continuum region and the
other in the line region. A Student's $t$ test was performed in order
to evaluate whether these two values are significantly different:
if $t > 2.6$ in our data the significance level is less than 0.01,
i.e. the difference is highly significant. The results are 
reported in Table 1, and the ratios of the two standard
deviations are shown in Fig. 1. The error bars are actually 
overestimated, since they reflect the nonuniform variability 
within the line profile. The most extreme case is the binary 
HR 6681, which has very large variations confined just in the 
line wings.

\section{Results and discussion}
In this paper we present the relevant results for our
scientific motivation, that is those for HR 6139 and HR 5028.
A more detailed presentation and discussion of the results for
all the stars along with comparisons with possible models will
be reported in another paper. Here we just mention that the
variability of HR 5925 and HR 6681 can be easily explained by
their (previously unknown) spectroscopic binariety.
%-------------------------------------------------------------
   \begin{figure}
   \centering
   \includegraphics[bb=120 150 500 720, width=8cm]{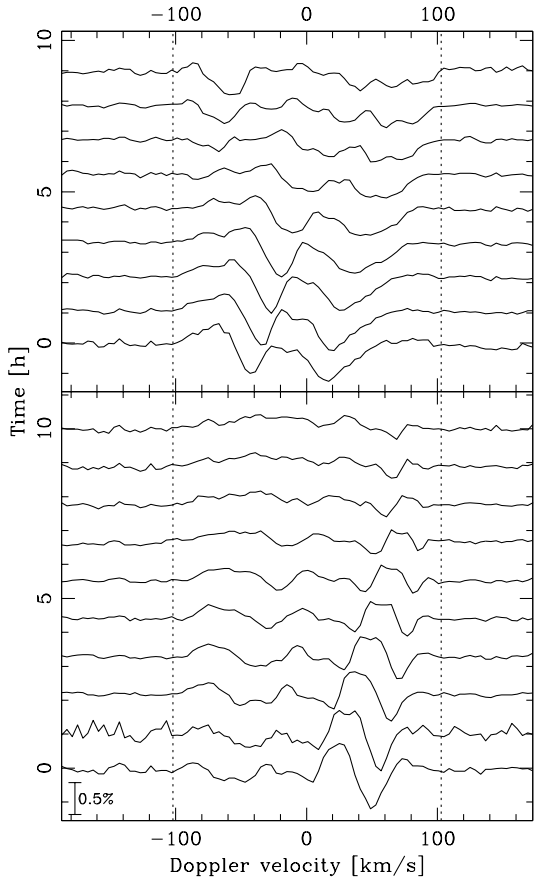}
      \caption{Variations of the mean line profile of HR 6139 during
      two consecutive nights; the average mean line profile has 
      been subtracted. The dashed lines correspond to the 
      borders of the average mean line. The progression of the
      shape of the profile suggests the presence of nonradial
      modes with a variability timescale of many hours. The bar 
      in the lower left corner indicates an amplitude of 0.5\%
      of the continuum flux.
              }
         \label{Fig2}
   \end{figure}
%-------------------------------------------------------------
%-------------------------------------------------------------
   \begin{figure}
   \centering
   \includegraphics[bb=120 340 500 720, width=6.2cm]{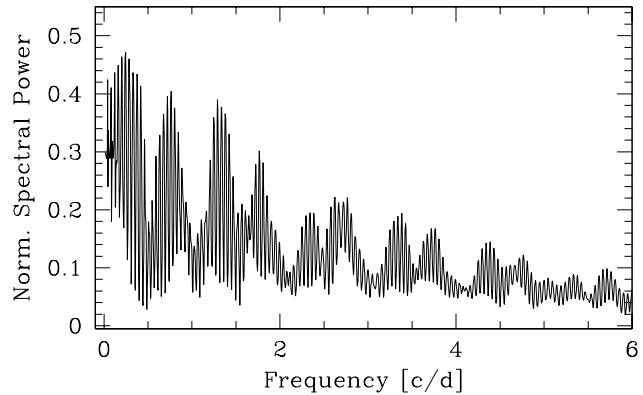}
      \caption{Global least-squares power spectrum 
   (Mantegazza, \cite{mant1}) of the mean line profile time series
   of HR 6139. This spectrum takes into account the information 
   about the variability of the whole mean line profile. One should
   note the large power at low frequencies. 
              }
         \label{Fig3}
   \end{figure}
%-------------------------------------------------------------

Fig. 2 shows the line profile variations of HR 6139 during two 
consecutive nights. The presence of nonradial oscillations propagating in the
sense of the stellar rotation is evident. There is a progression
of the complex shape of the mean line profile, and the timescale of the
variability suggested by this progression is of many hours. 
Fig. 3 shows the ``global least-squares power spectrum'', 
that is the result of the pixel-by-pixel frequency analysis
(Mantegazza, \cite{mant1}). 
This spectrum takes into account the information about the variability
of the whole line profile. It shows that the spectral power of the 
line profile variations  is concentrated at frequencies considerably
lower than 5 c/d, which is the estimated value of the
fundamental radial mode frequency of HR 6139. Therefore the
line profile variation is probably due to prograde $g$-modes.
The spectral window is too much entangled to allow the discrimination
of the individual modes.  HR 6139 is a known small amplitude 
photometric variable ($\sim 0.02$ mag), and Koen (\cite{koe}), using HIPPARCOS data, 
detected two periods, 0.38 and 0.64 c/d. However, the two photometric periods
are not sufficient to explain all the line profile variations.
The estimated $T_{\rm{eff}}$ from $uvby{\beta}$ and Geneva photometry is about 8800 K, 
and the absolute magnitude derived from HIPPARCOS parallax, 
with correction for a small reddening $E(b-y)=0.006$, is $M_V =-0.30\pm0.21$; there
is no evidence of binariety, and HR 6139 looks like a normal single star.
On the whole, its variability reminds of that of the
SPB stars (e.g. DeCat et al., \cite{dec}; Aerts \& Kolenberg,
\cite{aer}). However the star is much cooler than the
cool border of the SPB star instability strip predicted by the models, even
if we take into account the proposed mechanisms that can
extend the strip. On the other hand, it is hotter than the blue border
of the $\delta$ Scuti star instability strip. Therefore HR 6139
is apparently challenging the present theoretical models.
%-------------------------------------------------------------
   \begin{figure}
   \centering
   \includegraphics[bb=120 130 500 570, width=6.2cm]{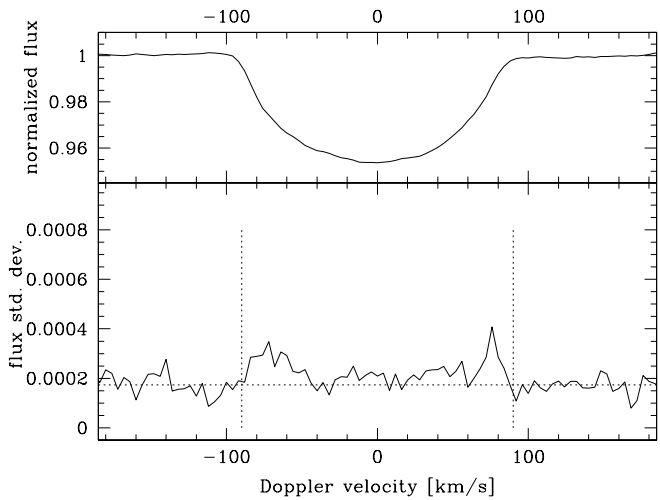}
      \caption{Upper panel: average line profile of HR 5028. Lower
    panel: the tiny variability indicated by the relatively large
    standard deviation from the mean of the average line in comparison 
    with the relatively small standard deviation of the adjacent
    continuum. The horizontal dashed line represents the mean
    level of the standard deviation of the continuum, which can be 
    interpreted as noise level.
              }
         \label{Fig4}
   \end{figure}
%-------------------------------------------------------------
%-------------------------------------------------------------
   \begin{figure}
   \centering
   \includegraphics[bb=120 130 500 570, width=6.2cm]{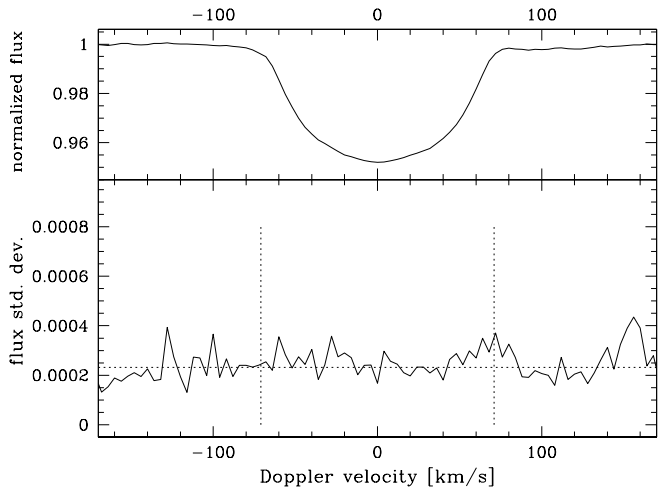}
      \caption{Average line profile of HR 5670. In this 
    case the standard deviation from the mean of the 
    average line is not different from that of the mean level
    continuum (dashed line).
              }
         \label{Fig5}
   \end{figure}
%-------------------------------------------------------------

It is too early to identify HR 6139 as a prototype star for the
mechanism found by Browning et al. (\cite{bro}), and we suspect that
such a mechanism, if present, should give smaller effects. 
In this respect, the result for HR 5028 shown in Fig. 4 could be of 
some interest. In the lower panel of this figure we compare the 
behaviour of the standard deviations of the velocity bin
timeseries in the line profile with those in the adjacent continuum 
regions. The dashed horizontal line is the average standard deviation 
in the continuum, and it can be interpreted as the noise level. 
The standard deviations of the timeseries of the line profile bins 
are slightly larger. As a reference, in the upper panel we show the average 
line profile. In Fig. 5 it is shown for comparison purposes the profile
of the constant star HR 5670; in this case there are no differences
between the line profile and the adjacent continuum.  
We are not able to offer an explanation of the {\em tiny} variability of the
line profile bins with respect to the continuum of the stars
such as HR 5028, and for the present
we cannot ascribe it to possible very small oscillations. A more
solid database is needed to address this issue adequately.

We intend to continue the survey in order to find other cases
of $g$-mode (and possible $p$-mode) pulsators among early-A type
stars. In the past fifty years there were several claims of (or 
attempts to verify) the existence of pulsating stars in this narrow 
spectral region, hence it is mandatory to increase the observational 
evidences before attempting to define a new possible class.

\begin{acknowledgements}
The authors gratefully acknowledge financial support from MIUR 
for grant PRIN 2004 ``Asterosismologia'' (PI: L. Patern\`o). 
A.M. acknowledges financial 
support also from the Prodex-ESA Contract 15448/01/NL/Sfe(IC).
\end{acknowledgements}

\end{document}